\documentclass[conference]{IEEEtran}
\IEEEoverridecommandlockouts
\usepackage{cite}
\usepackage[T1]{fontenc}
\usepackage{amsmath,amssymb,amsfonts}
\usepackage{algorithmic}
\usepackage{booktabs}
\usepackage{graphicx}
\usepackage{caption}
\usepackage{tabularx}
\usepackage{float}
\usepackage{textcomp}
\usepackage{placeins}
\usepackage{xcolor}
\graphicspath{{figures/}}

\def\BibTeX{{\rm B\kern-.05em{\sc i\kern-.025em b}\kern-.08em
    T\kern-.1667em\lower.7ex\hbox{E}\kern-.125emX}}
\begin{document}

\title{
Tweaking Metasploit to Evade\\Encrypted C2 Traffic Detection
}

\author{
\IEEEauthorblockN{Gonçalo Xavier, Carlos Novo, Ricardo Morla}
\IEEEauthorblockA{\textit{FEUP and INESC TEC, } 
\textit{University of Porto}\\
Porto, Portugal \\
\{ up201604506, carlos.novo, ricardo.morla \} @ fe.up.pt}
}
\maketitle

\begin{abstract}
    Command and Control (C2) communication is a key component of any structured cyber-attack.
    As such, security operations actively try to detect this type of
    communication in their networks.  This poses a problem for legitimate
    pentesters that try to remain undetected, since commonly used pentesting
    tools, such as Metasploit, generate constant traffic patterns that are
    easily distinguishable from regular web traffic.
    In this paper we start with these identifiable patterns in 
    Metasploit's C2 traffic and show that a machine learning-based
    detector is able to detect the presence of such traffic with high accuracy,
    even when encrypted.  
    We then outline and implement a set of modifications to the Metasploit
    framework in order to decrease the detection rates of such classifier. 
    To evaluate the performance of these modifications, we use two threat models with increasing awareness of
    these modifications. We look at the detection evasion performance and at the byte count and runtime overhead of the modifications. 
    Our results show that for the second, increased-awareness threat model the framework-side traffic modifications yield a better detection avoidance rate (90\%) than payload-side only modifications (50\%). We also show that although the modifications use up to 3 times more TLS payload bytes than the original, the runtime does not significantly change and the total number of bytes (including TLS payload) reduces. 

\end{abstract}


\section{Introduction}

Pentesters put themselves in the shoes of attackers~\cite{hacking-ethnogr} to test the security of
their client's systems. Remote control of compromised assets plays a major role
in pentesting; this is achieved through the deployment of payloads and the
communication between the payloads and the pentester's command and control
software~\cite{metasploit-payload}. Detecting this communication is important for security operations,
not as much to detect legitimate pentesters as to detect actual attackers that
use pentesting tools~\cite{metasploit-study}. This poses a challenge to the legitimate pentester -- if
C2 communications are blocked how can the pentester uncover vulnerabilities and
other security issues elsewhere in the target system? IP-specific rules for the
pentester could be enacted by the client's IT staff to allow the pentester's
traffic, but this goes against the premise that a pentest is as good as the
assumptions for the attacks are weak. Clients may not take seriously the
security issues that the pentester finds if they think the attack is only
possible because they allowed it. One way for the pentester to overcome these
difficulties is to  modify its C2 traffic in order to avoid detection.

In this paper we explore different ways in which the pentester can modify its traffic and what is the impact of those modifications in evading a detector.
Pentesting C2 traffic can be distinguished from other, non-C2 traffic with
machine learning techniques. Machine learning is especially relevant if the C2
traffic is encrypted and blacklisting server IPs or certificates is not viable
because of inexpensive certificate replacement or cloud-based server-side IP
sharing~\cite{ipvest}. Different pentesting frameworks are likely to have different C2
traffic profiles that a machine learning-based detector may need to learn to
distinguish from normal traffic. An approach for C2 traffic communication is to
have the payload periodically open a connection to the server, inquiring for
orders or reporting results. This ends up creating a regular traffic pattern
that can be much different from web and other, non-C2 traffic -- and makes
it relatively easy to detect. This is the case of Metasploit\footnote{https://github.com/rapid7/metasploit-framework/}, one of the best
known pentesting tools. Armed with traffic captures of C2 traffic from a
specific tool like Metasploit and of non-C2 traffic from their local networks,
security operations are able to train a machine-learning based detector of C2
traffic for their networks. In section \ref{sec:MetasploitC2Detector} we show results confirming that
for a specific dataset with Metasploit traffic it is possible to train a
detector with extremely high performance.

Knowing that its C2 traffic is regular and may be easily detected, the
pentester may feel the need to change the traffic pattern to evade a detector. In section \ref{sec:EvadingRegularMetasploitDetector} we describe a set of traffic pattern changes to C2 traffic and show how the Metasploit
Framework can be modified in order to generate such traffic and evade a machine
learning-based detector.
We then consider a second threat model in section \ref{sec:EvadingAwareDetector}, where the target's network detection
system is aware of the modifications made to the C2 traffic and the pentester employs adversarial
learning techniques~\cite{sok-sp-ml} in order to evade the detector.
We present detector evasion results and discuss the overhead and performance impacts associated with the framework modifications.

\section{Related Work}

The communication patterns of open-source
C2 frameworks have been studied before.
For example, \cite{kn:InDepthStudyOpenSourceC2} highlights common practices and
communication behaviours present in some of the most popular pentesting
frameworks, and shows how a machine learning model could be trained based on
these behaviours to distinguish each framework's traffic. 
Additionally, and specifically focusing on Metasploit, 
\cite{kn:AnalysisMeterpreterPostExploitation} studied the entire behaviour of a
Metasploit payload inside a victim's host, focusing on 
C2 communication patterns and on the presence of the  payload 
in the victim's host memory.
Although these studies are important for improving the performance of
intrusion detection systems (IDS's), they fail to inform a legitimate
pentester on how to overcome such identifiable characteristics and evade
detection systems.
Additionally, due to the active and open-source development of such
frameworks, these studies have quickly become outdated and no longer accurately
describe the current implementation of these tools.

Evading detection systems has been a concern for the Metasploit's development
team, as defenders more quickly react to new threats by automatically analyzing and updating detection rules~\cite{wang2013metasymploit}.
In 2018, the Metasploit team incorporated evasion
techniques like code obfuscation, code stub injection and encryption in the
Metasploit Framework modules\cite{kn:MetasploitAVEvasion}. 
Casey et al. \cite{kn:ComparativeEvasionModule} studied the performance of these
modules in evading commonly used anti-virus software and showed that
most anti-viruses are successful at detecting them, likely due to the public availability of the documentation of these modules. 
The techniques used in these evasion modules are targeted at evading
static signature-based antivirus checks like binary file analysis
and they don't change the C2 patterns that can give away the presence of
pentesting tool traffic on the network.

Other studies have focused on the use of adversarial learning techniques
against machine learning-based detection systems, specially targeting 
malware's traffic characteristics.
For example, \cite{kn:GenerativeAdversarialAttacksAgainstIDS} showed how a
generative adversarial network (GAN) could be used to adapt the inter-packet
time and packet length variance of botnet attacks, in order to bypass an IDS
model.
Based on the same type of traffic features, Yang et al.
\cite{kn:AdversarialExamplesAgainstIDS} showed the performance of different
adversarial algorithms against an IDS, in a black-box scenario
and the authors of \cite{kn:tiki-taka} have proposed multiple defense methods against adversarial attacks targeting intrusion detection systems. 
In a more C2-specific work, Rigaki et al. \cite{kn:Rigaki2018BringingAG}
showed how by adapting a Remote Access Trojan (RAT) C2 communication scheme,
according to a GAN, it was possible to mimic the traffic of a legitimate
application, thus evading a detector.
Although these studies contribute to highlighting the weakness of IDSs against
adversarial examples, they all focus on incorporating adversarial learning in
malware samples, not on pentesting tools.
Specifically, these studies don't focus on  adapting source code such that an IDS that focuses on encrypted traffic can be evaded.

Another conceptually similar topic is the one of censorship circumvention, where an actor masks traffic features by mimicking allowed traffic in an attempt to bypass the censor's detector. The authors of \cite{kn:dynamic-traffic-camouflaging} have used a GAN for automatic traffic generation and censor circumvention, and advancements made in this research field may generally be used by pentesters and vice versa.



\section{Metasploit C2 Traffic Patterns}
\label{sec:MetasploitC2TrafficPattern}

During a normal Metasploit exploit process the deployed payload and the Metasploit framework have to communicate.
When using a stageless payload this communication is bootstrapped as follows:
\begin{enumerate}
    \item The exploit occurs and the payload starts on the victim's host;
    \item The payload's dispatch loop starts and the payload
        reaches out to the framework on a specific host and HTTP
        URL; 
    \item The Metasploit framework on the pentester host recognizes this request as a new session and
        negotiates TLV C2 encryption keys. Afterwards, any communication
        for this session between the payload and the Metasploit
        Framework consists of GET or POST requests with encrypted data.
\end{enumerate}

\begin{figure}[h!]
    \centering
    \includegraphics[width=1\linewidth,keepaspectratio]{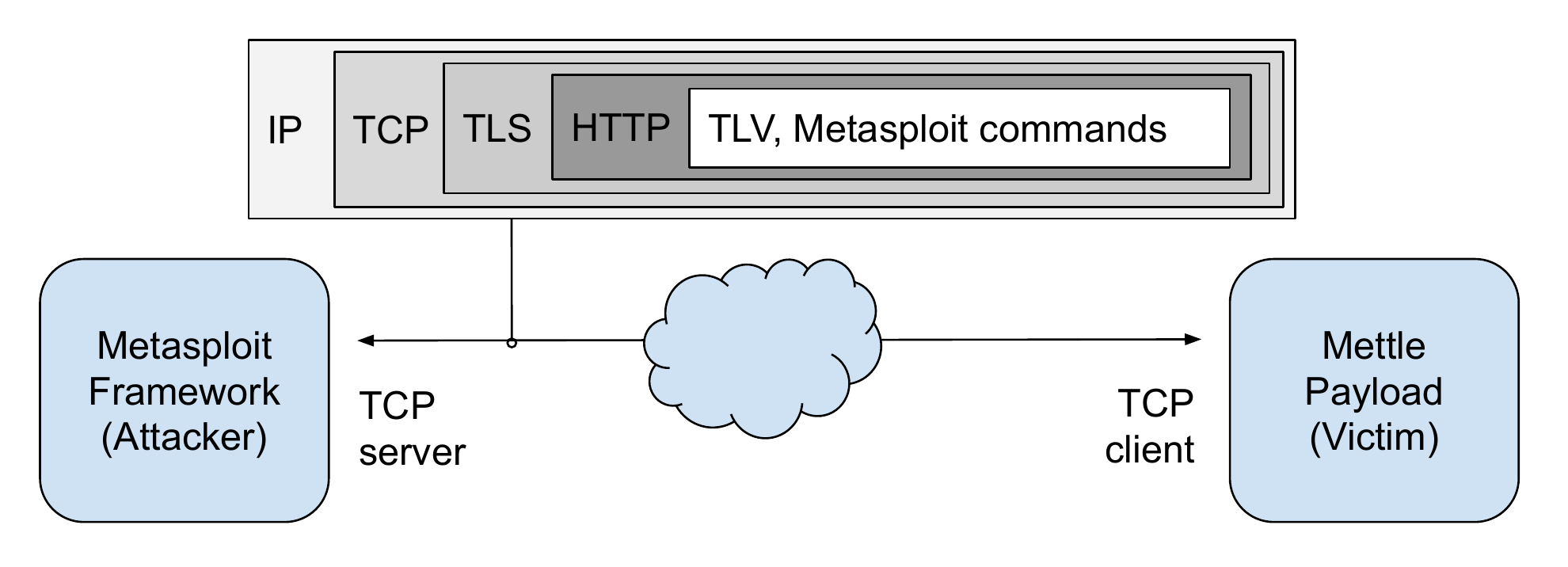}
    \caption{Metasploit's packet structure when using a HTTPS payload. }
    \label{fig:PacketStructure}
\end{figure}

In our setup we configure the Mettle payload (Mettle is the C-based Metasploit payload\footnote{https://github.com/rapid7/mettle}) to use HTTPS (HTTP over TLS), which additionally secures the bootstrapping and prevents direct eavesdropping of e.g. the type and headers of the HTTP request that the payload issues. 
When using this type of payload, the Metasploit framework communicates using
packets that follow the structure presented in Figure
\ref{fig:PacketStructure}, with the encrypted data transmitted between the
Metasploit Framework and the payload encapsulated in the TLV layer of each
packet.
As the Metasploit framework does not
internally differentiate between HTTP and HTTPS traffic, its communication
workflow is the same whether using HTTP or HTTPS.

\FloatBarrier

\begin{figure}[h!]
    \centering
    \includegraphics[width=0.6\linewidth,keepaspectratio]{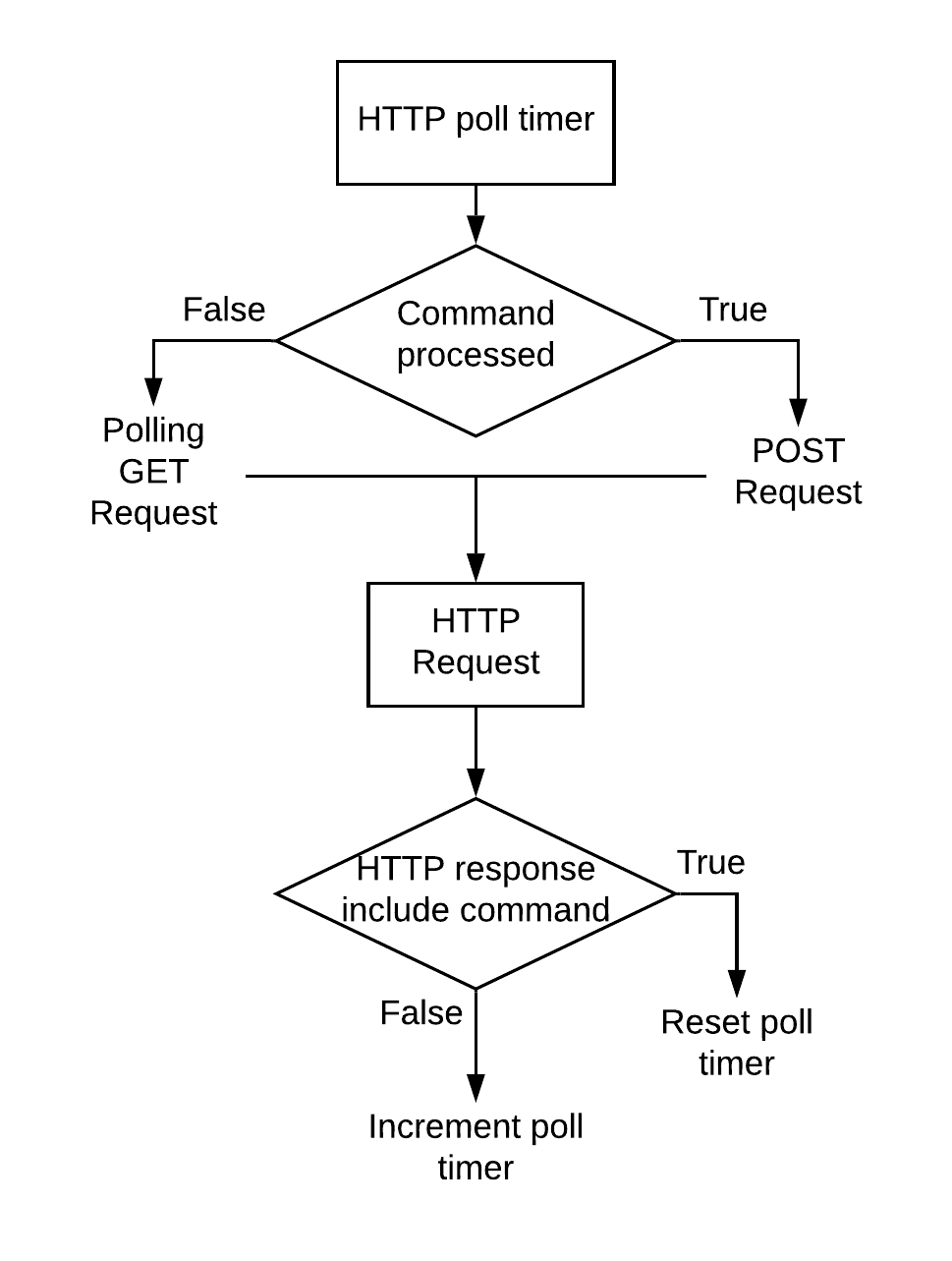}
    \caption{Mettle HTTP communication scheme}
    \label{fig:MettleHTTPCommSchemeSimple}
\end{figure}

Figure \ref{fig:MettleHTTPCommSchemeSimple} shows the payload's communication workflow after a secure communication session has been established.
The payload's
dispatch loop repeatedly contacts the framework in a pooling manner, using
empty HTTP GET requests on a new URL generated by the framework. The framework responds to each of these GET requests by sending any queued commands
back to the payload. These commands are then individually processed and their results
returned using POST requests back to the framework.
If no commands are returned to the payload, the interval between the pooling
requests doubles up to a maximum of 10 seconds.

\begin{figure}[h!]
    \centering
    \includegraphics[ width=0.7\linewidth,keepaspectratio]{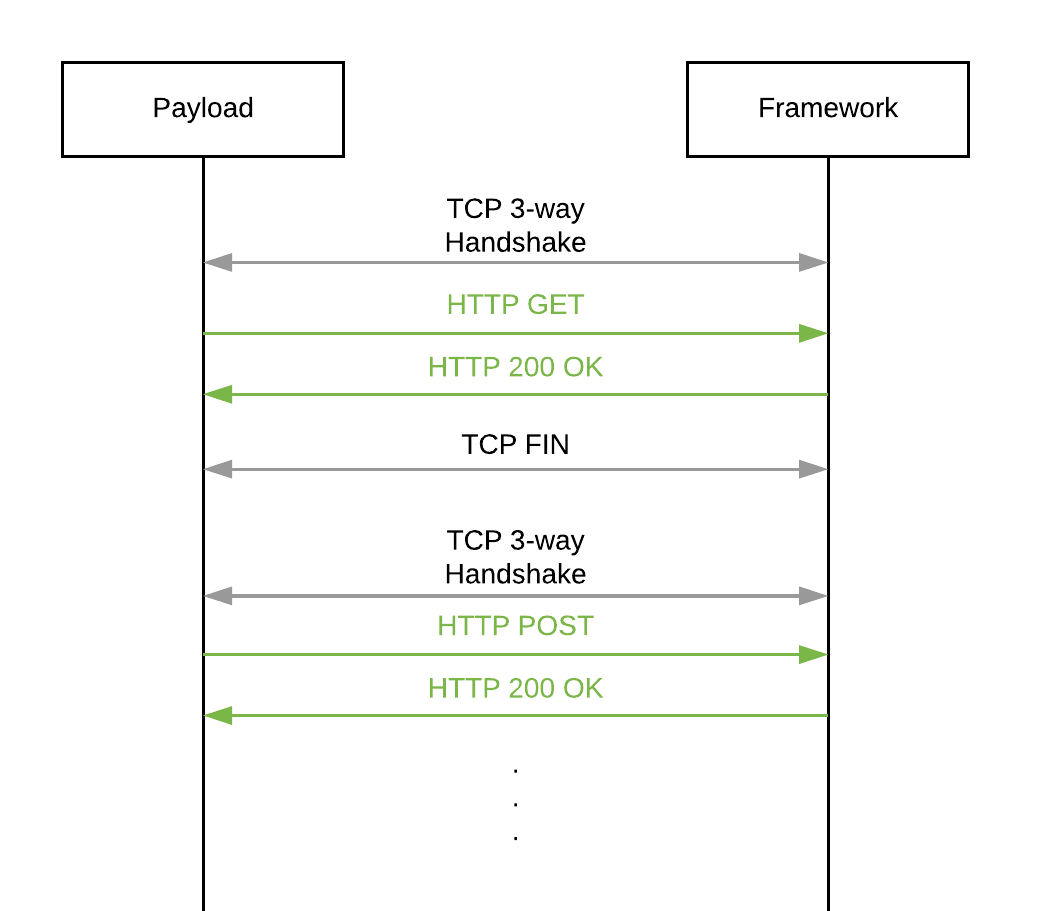}
    \caption{Diagram for a traffic flow sample between HTTP payload and the Metasploit framework}
    \label{fig:HTTPGETFlow1}
\end{figure}

\begin{figure}[h!]
    \centering
    \includegraphics[ width=1\linewidth,keepaspectratio]{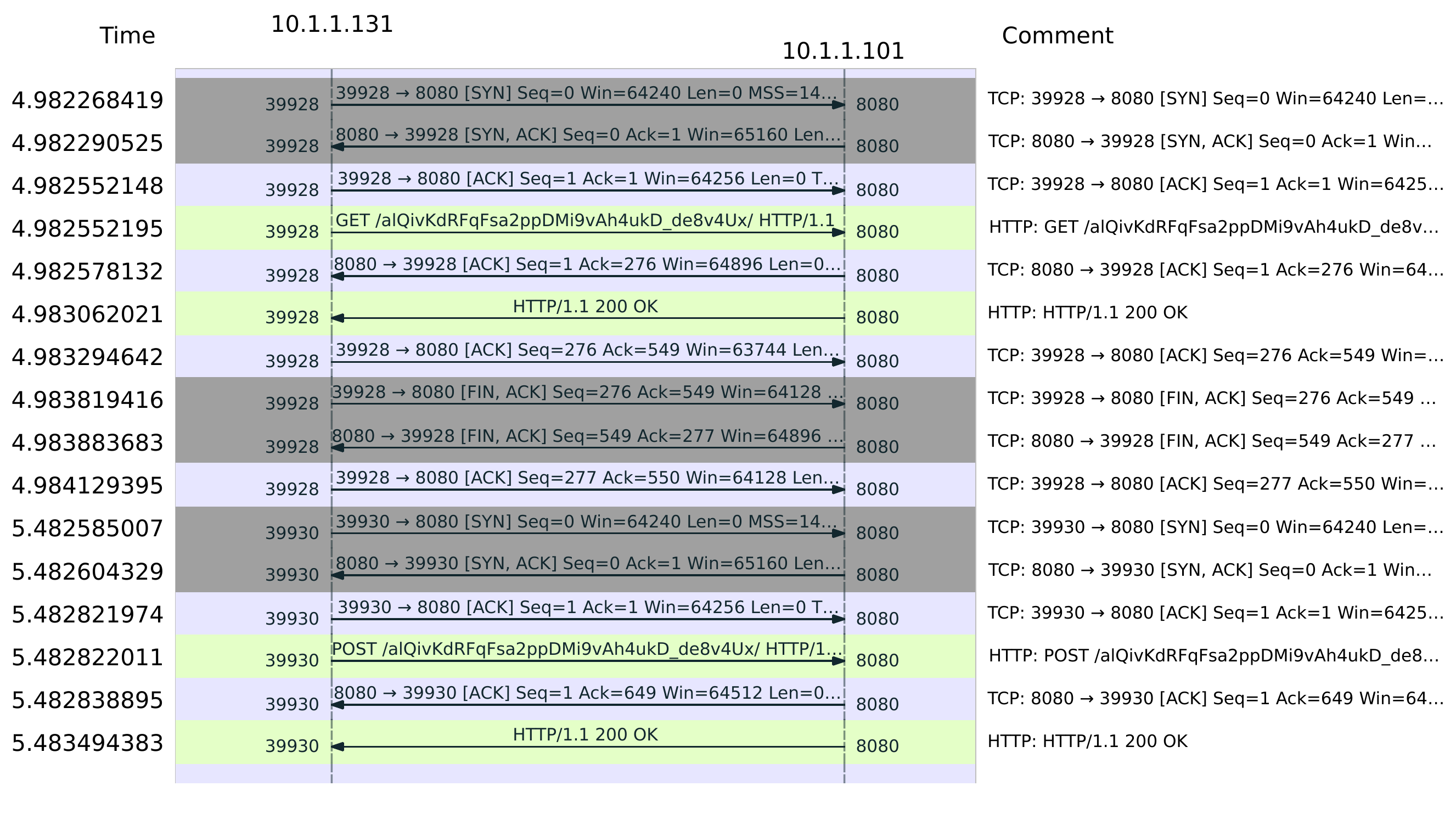}
    \caption{Wireshark capture for a traffic flow sample between HTTP payload and the Metasploit framework}
    \label{fig:HTTPGETFlow2}
\end{figure}

This pooling pattern generates very consistent and periodic traffic 
not commonly present in regular web traffic. Mettle
uses the \textit{libcurl} library to execute its HTTP requests. \textit{curl} configures requests including custom options, headers, and payload through the \textit{handle} programming construct. Every time a requests is made and its response is received, Mettle terminates
the corresponding \textit{handle}.
This automatically causes a new TCP connection to be initiated and terminated whenever the 
payload needs to communicate with the framework. This can be seen in Figures
\ref{fig:HTTPGETFlow1} and \ref{fig:HTTPGETFlow2}.

\FloatBarrier

\FloatBarrier

\FloatBarrier

As a consequence of this behaviour, when using an HTTPS payload, 
the traffic flow between Metasploit and the payload rarely exceeds more than two TLS Application Data records (type 23, AppData).
This stands out when
compared to non-C2, web traffic.
Table \ref{fig:BTC2TLSFlow} highlights this characteristic 
by grouping all TLS records belonging to the same TCP
connection and then filtering them by their TLS record type. 
C2 traffic only shows 1 HTTP request per connection, which
produces 2 TLS AppData records -- one in each direction. Web traffic shows several TLS AppData records exchanged
during the same TCP connection.

\begin{table}[h]
\begin{center}
{\scriptsize

\begin{tabular}{rrrrrrrrr} 
\toprule\textbf{1} & \textbf{2}& \textbf{3}& \textbf{4}& \textbf{5}& \textbf{6}& \textbf{7}& \textbf{8}& \textbf{9}\\ 
\midrule
 396 & 5698 & 454 & 3552 & 452 & 16408 & 16408 & 16408 & 16408\\
 32 & 560 & 1280 & 32 & 528 & 512 & 32 & 560 & 1280\\
 448 & 1494 & 453 & 2644 & 452 & 16408 & 16408 & 16408 & 16408\\
 32 & 560 & 1280 & 32 & 528 & 512 & 32 & 528 & 512\\
 \midrule
 288 & 288 &  &  &  &  &  &  & \\
 704 & 176 &  &  &  &  &  &  & \\
 288 & 288 &  &  &  &  &  &  & \\
 624 & 176 &  &  &  &  &  &  & \\
 \bottomrule 
 \end{tabular}

}

    \caption{First 9 TLS Application Data record sizes for C2 traffic and non-C2 (web) traffic for sample TCP connections. Top: non-C2 traffic, bottom: C2 traffic.}
    \label{fig:BTC2TLSFlow}
\end{center}
\end{table}

The communication workflow we describe in this section is for the standard Metasploit
and Mettle versions available online. 
In sections \ref{sec:EvadingRegularMetasploitDetector} and \ref{sec:EvadingAwareDetector} we show how to modify Mettle and Metasploit implementations to change traffic patterns while keeping the same workflow and evading a detector. 
Before that, in section \ref{sec:MetasploitC2Detector}, we show how the standard communication pattern is easily detectable from non-C2 traffic without these modifications.

\section{Metasploit C2 Traffic Detection}
\label{sec:MetasploitC2Detector}

The model we use to detect C2 traffic takes as input a sequence with the sizes of the first 20 TLS AppData records in both directions of a TCP connection.  
These sequences of sizes are 
extracted from 
traffic captures using Cisco's Joy\footnote{
https://github.com/cisco/joy
} tool.
These features are then filtered by their TLS content type and the length of the
AppData records is fed to the model.  We pad flows with fewer than 20 TLS Appdata records using the value -1 and truncate flows with more than 20 TLS Appdata records.

Internally, the model consists of a multi-layer neural network with an input
layer of 20 nodes (for the first 20 TLS AppData records), 3 ``hidden layers"
of 2048, 1024 and 512 nodes respectively, all with a \textit{relu} activation
function, and a final output layer with 2 nodes for categorical classification with a \textit{softmax}
activation function.  
In each layer, we add a 20\% dropout layer to prevent
overfitting.
The \textit{softmax} function used in the output layer returns a 2-dimensional probability
distribution vector on whether the input sequence is considered a C2 traffic flow or
not.
The Adam optimizer was used for updating weights during training and we use categorical cross-entropy as the loss function.  

The model was trained and evaluated using a dataset of locally generated 
Metasploit C2 traffic together with a dataset of non-C2 traffic.  
Our locally generated Metasploit C2 traffic consists of a series of Metasploit commands typically used during a pentest, such as
\texttt{sysinfo},
\texttt{ps},
\texttt{ipconfig},
\texttt{route},
\texttt{download /etc/shadow},
\texttt{ls}, 
that are invoked automatically once the payload contacts Metasploit. We repeatedly spawn payloads on a virtual machine and capture the traffic the payloads generate towards the Metasploit host. Our non-C2 traffic was generated using BrowserTime\footnote{
https://github.com/sitespeedio/browsertime
} and consists of 10 visits to each of the top 1k Alexa web sites\footnote{
http://s3.amazonaws.com/alexa-static/top-1m.csv.zip
}.

We achieved a detection accuracy of 99\% on 111k balanced C2 and non-C2 samples.  
This high accuracy confirms our intuition that Metasploit's C2 traffic is highly identifiable and that a pentester could be easily detected when using this framework.

\section{Evading 
a Regular 
Metasploit Detector }
\label{sec:EvadingRegularMetasploitDetector}

The detector model developed in section \ref{sec:MetasploitC2Detector} 
is tested with Metasploit C2 samples generated from the standard Metasploit framework.
Our first new threat model considers that a pentester is smart enough to
attempt to evade such a detector by modifying its C2 traffic.

\textit{Threat model \#1:} 
A C2 traffic detector, aware of the traffic profiles generated by
regular web traffic and the standard Metasploit framework, is able to
consistently detect the presence C2 traffic.
Knowing about this, and to remain undetected, a pentester is able to modify the communication scheme of
the Metasploit framework in two ways: 1) increase the size of TLS records (section \ref{subsec:PacketStuffing}), and 2) change the number of HTTP requests used within the same TCP connection (section
\ref{subsec:HTTPRequestPerTCP}). We describe both approaches and their evasion results next.

\subsection{Packet Stuffing}
\label{subsec:PacketStuffing}

\begin{figure}[h!]
    \centering
    \includegraphics[width=1\linewidth,keepaspectratio]{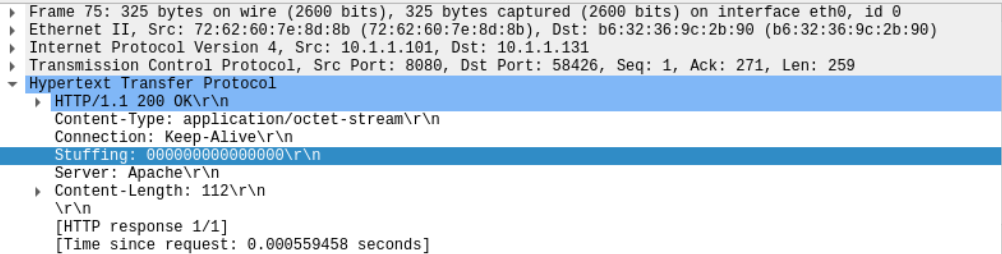}
    \caption{HTTP header used in packet stuffing}
    \label{fig:HTTPStuffHeader}
\end{figure}

A noticeable aspect of C2 traffic is the small size of TLS records when compared
to web traffic. To try and change this characteristic, we develop a generic packet stuffing method. This method operates on the ``HTTP Request" block of the Mettle communication
scheme (as described in Figure \ref{fig:MettleHTTPCommSchemeSimple}) and adds a
stuffing HTTP header to each HTTP request. 
This has the indirect effect of increasing the size of the TLS records analysed by the detector's model. It also does not change the functionality of the framework.
Figure \ref{fig:HTTPStuffHeader} shows how this stuffing occurs on actual packets.
By applying a similar method on the framework-side it is possible to
achieve packet stuffing in both sides of the communication.

To evaluate the evasion rate of this technique using the same methodology as
in section \ref{sec:MetasploitC2Detector}, we generated 
two datasets of C2 traffic,
one with constant packet stuffing of 50 bytes
and one with random size packet stuffing.
In both datasets packet stuffing occurs in the packets from the framework and from the payload.
We then assess the datasets' evasion rate.
We define evasion rate as the percentage of C2 traffic samples that were misclassified by the detector. Table \ref{table:packetStuffing} shows the evasion rate for regular Metasploit traffic and for 50 byte and random stuffing traffic.

\FloatBarrier

\begin{table}[h]
\begin{center}
        \begin{tabular}{lcc}
            \toprule
            \textbf{Dataset } &  \textbf{Total number} &
            \textbf{Evasion }  \\
            \textbf{type} &  \textbf{of samples} &
            \textbf{ rate}  \\
            \midrule
            Regular Metasploit &     111853 &  9 (0.01\%) \\
            \\
            50 byte Stuffing  &     67513 &  3259 (4.83\%) \\
            \\
            Random byte size Stuffing  &     66768 &  66680 (99.87\%)\\
            \bottomrule
        \end{tabular}
        \caption{Metasploit C2 evasion rate results 
        (default vs. packet stuffing)}
        \label{table:packetStuffing}
\end{center}
\end{table}

\FloatBarrier

After implementing this technique,
when using 50 byte length headers as stuffing,
the pentester would only be able to achieve
a slightly better evasion rate as when using the unchanged Metasploit tool.  
However, when using random length headers, the pentester can almost completely
evade the detector.
Although this technique alone might be enough to trick a detector aware of the
traffic generated by the standard Metasploit Framework, it does not alter
the other identifiable characteristic
of the Metasploit framework: the number of HTTP requests per TCP connection; we consider this next.

\subsection{HTTP Request per TCP Connection}
\label{subsec:HTTPRequestPerTCP}
\FloatBarrier

As shown in Figures \ref{fig:HTTPGETFlow1} and \ref{fig:HTTPGETFlow2}, in Table \ref{fig:BTC2TLSFlow}, and as
discussed in section \ref{sec:MetasploitC2TrafficPattern}, the immediate
termination of the \textit{curl} 
\textit{handle} causes very short TLS Application Data
flows across all Metasploit traffic. 
This could be seen as the most distinguishable characteristic in Metasploit's
C2 traffic.
Several TLS AppData packets can be
included in the same TCP connection in order to make C2 traffic less distinguishable from web traffic, which typically has TCP connections with more TLS records.
To achieve this, the \textit{curl} \textit{handles} 
are now re-used until a 
number 
of Metasploit HTTP requests
have been exchanged in the same TCP connection.

\begin{figure}[h!]
    \centering
    \includegraphics[width=0.7\linewidth,keepaspectratio]{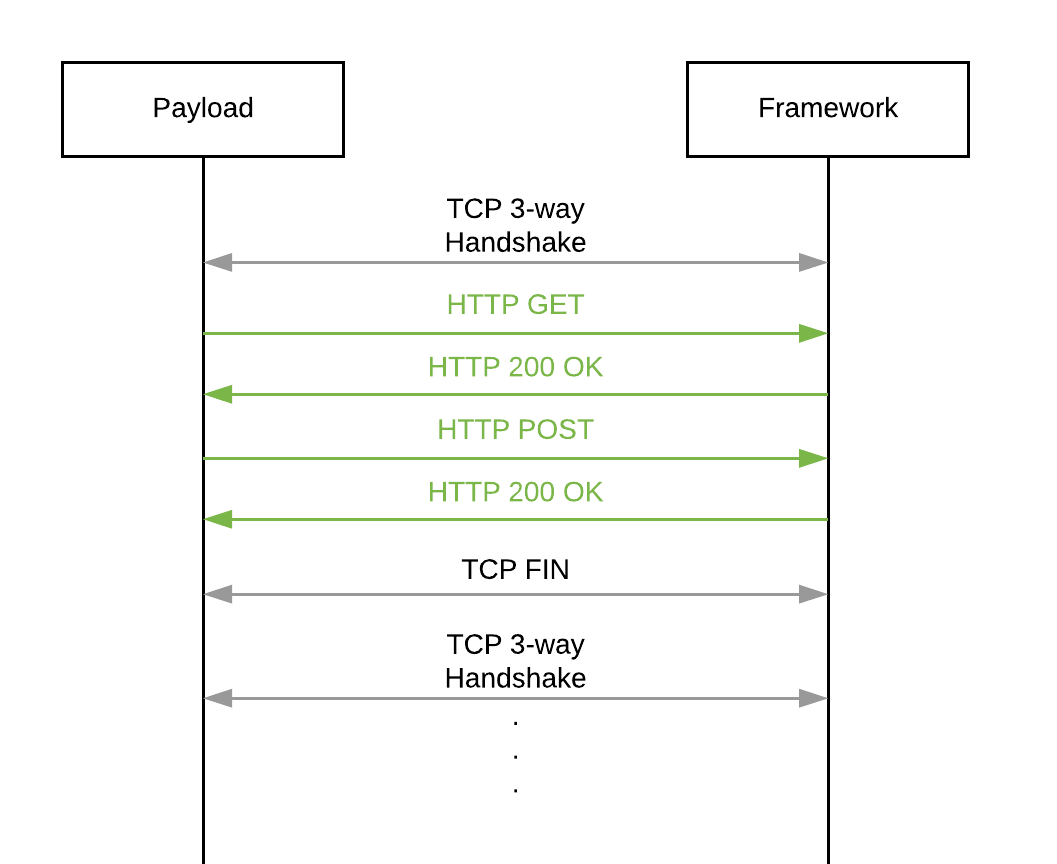}
    \caption{Diagram for a traffic flow sample between HTTP payload and the Metasploit framework with adaptation \ref{subsec:HTTPRequestPerTCP}. }
    \label{fig:MultipleReqFlow}
\end{figure}

Figure \ref{fig:MultipleReqFlow}  
shows the difference in Metasploit's TLS traffic flow with 
adaptation \ref{subsec:HTTPRequestPerTCP} when compared to Figure \ref{fig:HTTPGETFlow1}. We can now observe multiple HTTP requests in a single TCP connection. Although we changed the way HTTP requests are organized into TCP connections, the overall workflow remains the same and the GET and POST requests that can be observed correspond to polling and command response as in the default Metasploit payload.

Table \ref{fig:MultipleReqAppDataFlow}
shows the impact adaptation \ref{subsec:HTTPRequestPerTCP} has on the sequence of TLS AppData record sizes per TCP connection.
We observe this modification 
causes the Metasploit to generate a traffic profile that to a certain extent is closer to that of 
non-C2, web traffic shown in Table \ref{fig:BTC2TLSFlow}.

\FloatBarrier

\begin{table}[h]
\begin{center}
{\scriptsize

\begin{tabular}{rrrrrrrrr} 
\toprule\textbf{ 1} & \textbf{2}& \textbf{3}& \textbf{4}& \textbf{5}& \textbf{6}& \textbf{7}& \textbf{8}& \textbf{9}\\ 
\midrule
 336& 272& & & & & & & \\
 320& 112& 320& & & & & & \\
 320& 560& 768& 176& 784& 288& 560& 176& \\
 320& 304& 512& 176& 528& 304& 720& 176& \\
 320& 304& 1056& 176& 1072& 288& & & \\
 528& 176& 544& 288& & & & & \\
 576& 176& 592& 288& & & & & \\
 624& 176& & & & & & & \\
 \bottomrule 
 \end{tabular}

}
    \caption{First 9 TLS Application Data record sizes for sample TCP connections and C2 traffic using adaptation.}
    \label{fig:MultipleReqAppDataFlow}
\end{center}
\end{table}

We evaluate the evasion rate of
this technique using a C2 Metasploit dataset with 3 HTTP requests
per TCP connection and with a random number of HTTP requests per connection.
Table \ref{table:ReqTLSTCP} shows the evasion rate results for these techniques. By changing the amount of HTTP requests per TCP connection, a pentester can completely evade 
what was previously considered a reliable detector.

\begin{table}[h]
\begin{center}
    
        \begin{tabular}{lcc}
            \toprule
            \textbf{Dataset } &  \textbf{Total number} &
            \textbf{Evasion }  \\
            \textbf{type} &  \textbf{of samples} &
            \textbf{rate}  \\
            \midrule
            Default Metasploit &     111853 &  9 (0.01\%) \\
            \\
            3 HTTP requests  &     41094 &  41094 (100\%) \\
            per TCP connection&&\\
            \\
            Random \# HTTP requests  & 12415 &  12415 (100\%) \\
            per TCP connection &&\\

            \bottomrule
        \end{tabular}
        \caption{Evasion rate of multiple HTTP requests sent in the
        same TCP connection}
        \label{table:ReqTLSTCP}
\end{center}
\end{table}
\FloatBarrier

After noticing this clear improvement in his evasion rate, the pentester can
modify Metasploit to start using a random number of TLS AppData packets per TCP
connection, circumventing the periodic pattern that lead to the detection of C2 traffic in the first place.

\section{Evading an Increased Awareness Detector}
\label{sec:EvadingAwareDetector}

In order to further analyse the threat model presented in section
\ref{sec:EvadingRegularMetasploitDetector}, a more robust model, aware of the
previous changes made to the Metasploit's communication scheme, was trained 
and evaluated.
This model's layers and input features are the same as the one presented in 
section \ref{sec:MetasploitC2Detector}, however the training data now 
includes samples from the dataset ``Random \# HTTP requests per TCP Connection"
discussed in section \ref{subsec:HTTPRequestPerTCP}.
As a result of this addition, the detector is now able to detect Metasploit's
C2 traffic with multiple HTTP requests per TCP connection with an accuracy of
98.4\%. With this model in place, we consider a second threat model, as follows.

\textit{Threat model \#2:} 
A new C2 detector is trained with Metasploit traffic samples containing a random number of HTTP requests per TCP connection, as described in the beginning of this section.
Knowing about this, and to remain undetected, the pentester is able to additionally modify the communication pattern of the Metasploit framework from section \ref{sec:EvadingRegularMetasploitDetector}.B by adding stuffing bytes to the C2 traffic that are specifically chosen to mislead the detector through adversarial machine learning techniques.


Adversarial machine learning is a technique that tries to trick machine learning models into wrongly classifying specially generated inputs.
By taking advantage of the model's 
properties, these inputs only need to
slightly differ from normal ones in a specific way in order to be incorrectly
classified. In this paper, we use the Fast Gradient Sign Method~\cite{kn:FGSM}, an adversarial type of attack that changes the value of each input feature by a fixed $\epsilon$ in the direction of the detector's loss gradient,  $x^* = x + \epsilon sign(\nabla_x J(\theta,x,y))$, thus attempting to achieve a wrong classification.
This attack is implemented using the CleverHans \cite{kn:CleverHans} library. 

~\\
\subsection{Adversarial Packet Stuffing}
\label{subsec:AdversarialPacketStuffing}


By using the ``Random \# HTTP request per TCP Connection" dataset as the
original input for the CleverHans FGSM attack, we are able to generate a
sequence of TLS AppData packet lengths that is likely to trigger a non-C2
classification from the C2 traffic detector, while still following the profile
of C2 traffic.
By then combining the \textit{Packet Stuffing} and the \textit{Multiple HTTP
Request per TCP connection} modifications, it is possible to implement this
sequence in real C2 traffic. We add a number of stuffing bytes equal to the difference between the recommended adversarial size and the size of the HTTP request content; if the latter is larger than the former we do not add any stuffing to the HTTP request.

Adversarial packet stuffing can happen in two different scenarios:
\begin{itemize}
    \item \textit{One-side Adversarial Stuffing:}
        To keep the payload's size to a minimum and avoid 
        placing the entire adversarial sequence in a binary file inside the
        victim's host that may be subject to reverse engineering,
        the pentester only implements the
        adversarial sequence produced by CleverHans in the framework-side
        packets.
        In this case, the payload communicates normally, without any type of
        packet stuffing.
        This approach has the downside of only generating one half of the
        adversarial sequence (from framework to payload).
    \item \textit{Two-side Adversarial Stuffing:}
        While still trying to minimize the payload's size and not expose the
        full sequence of adversarial samples to eventual reverse engineering, the pentester modifies the
        framework-payload communication method to send the adversarial
        stuffing sizes to the payload. We describe this communication method in section \ref{subsec:StuffingProto}.
        This way, the entire adversarial sequence can be executed on both sides of the communication without encoding the sequence in the payload.

\end{itemize}
In both scenarios the number of TLS Application Data packets exchanged during
the same TCP connection is defined by the length of the adversarial
sequence used.

\subsection{Framework-to-payload Stuffing Protocol}
\label{subsec:StuffingProto}

\begin{figure}[h!]
    \centering
    \includegraphics[width=1\linewidth,keepaspectratio]{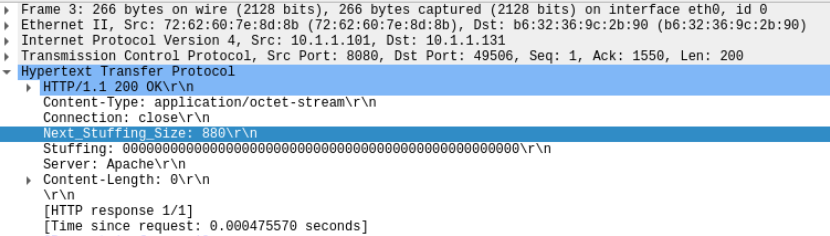}
    \caption{HTTP header used in framework-side packets during adversarial
    packet stuffing}
    \label{fig:AdvStuffHeader}
\end{figure}

Using a similar method as the one described in section
\ref{subsec:PacketStuffing}, 
we modified the Metasploit framework to 
include an HTTP header that informs the payload of what the next adversarial
stuffing size should be.
Using the received
adversarial size with the \textit{Packet Stuffing} method, the payload executes its 
part
of the adversarial stuffing sequence. Figure \ref{fig:AdvStuffHeader} shows a Wireshark capture with an example of the stuffing protocol header highlighted.

In the last framework-side request of an adversarial sequence,
the framework signals the
payload to terminate its \textit{curl handle} by changing the
\textit{Connection} HTTP header, from \textit{``Keep-alive"} to \textit{``close"}.
In that last request, the framework 
informs the payload of which value it should use for the stuffing size of the first TLS AppData record
in the next TCP connection. This way, the first value of new framework-payload TCP connections can also be part of an adversarial sequence.

\FloatBarrier

\subsection{Evasion Rate Results}

\begin{figure}[h!]
    \centering
    \includegraphics[width=0.9\linewidth,keepaspectratio]{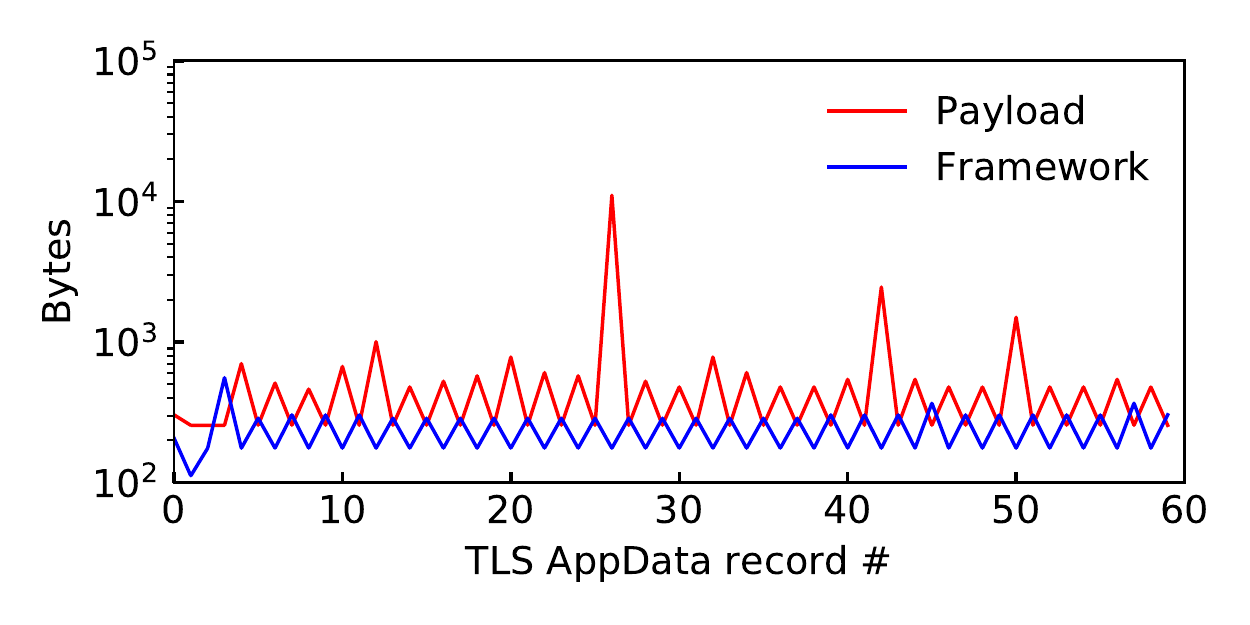}
    \caption{TLS AppData record sizes for a sequence of 12 different Metasploit commands, split by direction: from the payload and from the framework.}
    \label{fig:payload-framework}
\end{figure}

We evaluate the evasion rate of the adversarial packet stuffing technique against the C2 detector
of section \ref{sec:EvadingAwareDetector}
using a C2 Metasploit dataset with \textit{Adversarial Packet Stuffing}. Table \ref{table:adversarialPacketStuffing} shows our results. By incorporating
adversarial techniques into the Metasploit Framework, a pentester can significantly avoid detection even if 
the detector is aware of the modification techniques made to the original
Metasploit C2 traffic. We observe that when the technique includes adversarial stuffing from the framework side the detection avoidance rate is high at around 90\%, whereas if the adversarial stuffing is only executed at the payload side the avoidance drops to 50\%. This may be explained by the much more regular pattern of traffic going from the framework to the payload when compared with the traffic from the payload, as shown with an example in Figure \ref{fig:payload-framework}. Executing adversarial stuffing on the payload side in addition to adversarial stuffing on the framework side does not seem to significantly increase avoidance.

\begin{table}[h]
    \begin{center}
       \begin{tabular}{lcc}
            \toprule
            \textbf{Dataset } &  \textbf{Total number} &
            \textbf{Evasion }  \\
            \textbf{type} &  \textbf{of samples} &
            \textbf{rate}  \\
            \midrule
            Framework-side only &     14194 & 12976 (91.42\%) \\
            \\
            Payload-side only &     14986 & 7720 (51.51\%) \\
            \\
            Two-side  &     14185 & 12983 (91.53\%) \\
            \bottomrule
        \end{tabular}
        \caption{Metasploit C2 evasion rate results using adversarial 
        packet stuffing}
        \label{table:adversarialPacketStuffing}
    \end{center}
\end{table}
\FloatBarrier

\FloatBarrier

\begin{figure}[h!]
    \centering
    \includegraphics[width=0.9\linewidth,keepaspectratio]{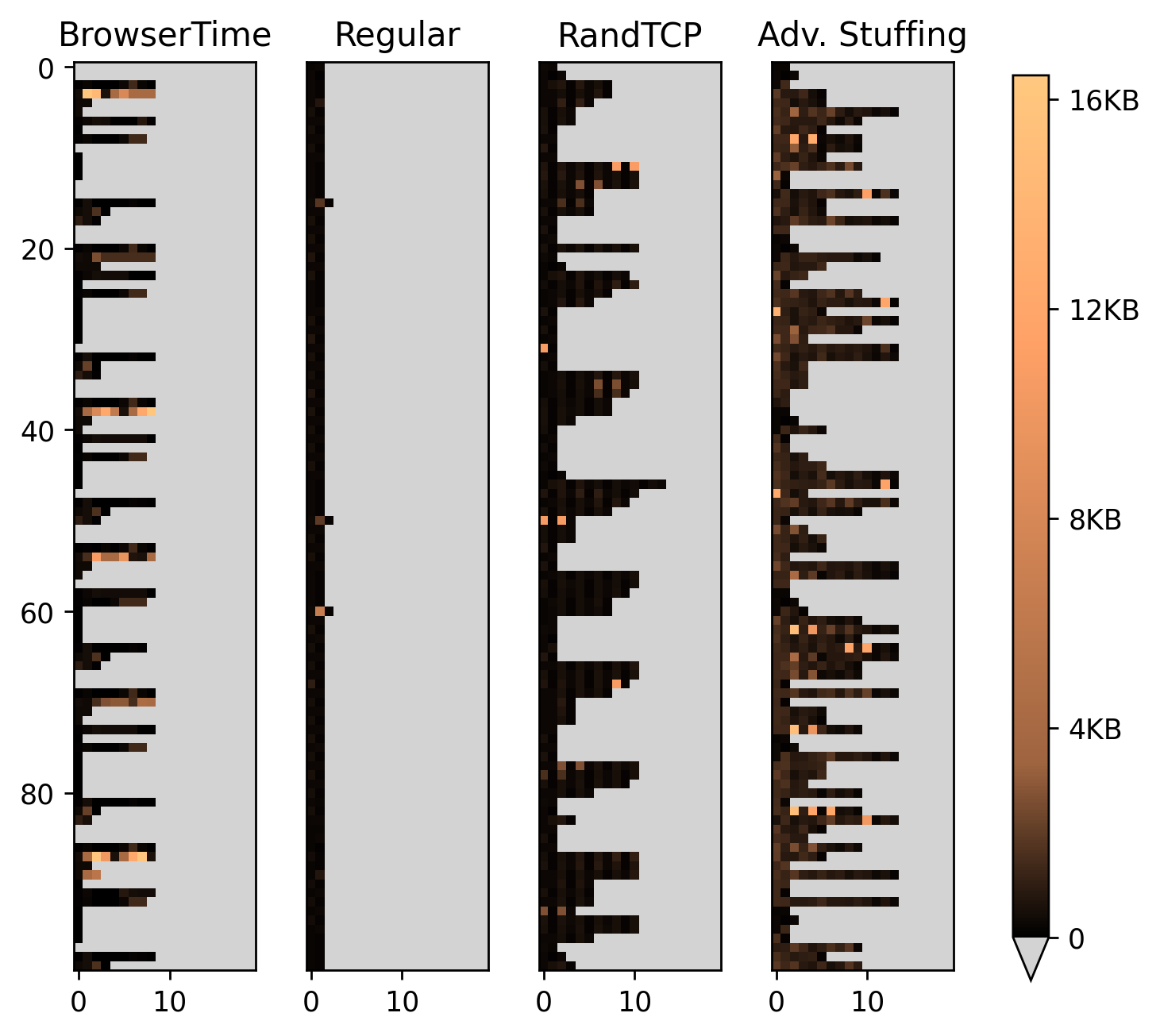}
    \caption{TLS traffic sample visualization: each line of pixels corresponds to a sequence of TLS AppData records for a different TCP connection;
    pixel colors represent the size of TLS AppData records. Left: BrowserTime web traffic, middle-left: original Metasploit C2 traffic, middle-right: Metasploit traffic with random number of HTTP requests per TCP connection; right: adversarial stuffing traffic.}
    \label{fig:traffic}
\end{figure}

Figure \ref{fig:traffic} helps us understand why our adversarial stuffing approach performs well against the increased awareness detector. The detector learns to distinguish BrowserTime web traffic (on the left column in Figure \ref{fig:traffic}) from a mix of regular traffic and traffic with a random number of HTTP requests per TCP connection (regular and RandTCP, on the two middle columns). 
We can observe that although the RandTCP traffic has more variability than the regular traffic, it is still far from looking like the web traffic. The adversarial stuffing approach takes advantage of this differences to evade the detector, which ends up deciding that most adversarial stuffing samples look more like web traffic than like the two middle columns, which is what it understands to be Metasploit traffic.
We also observe that adversarial stuffing and web traffic have similar TLS record sizes (most black, some yellow) yet adversarial stuffing has more TLS records per TCP flow than web traffic -- which, if used to retrain another detector, could be used to distinguish Metasploit from web traffic.

\subsection{Overhead}
\label{sec:Overhead}

In this section we look at the impact that our Metasploit modifications have on the number of bytes sent between framework and payload and on the time it takes to have the payload run a set of pentesting commands.

\FloatBarrier

\begin{figure}[h!]
    \centering
    \includegraphics[width=0.9\linewidth,keepaspectratio]{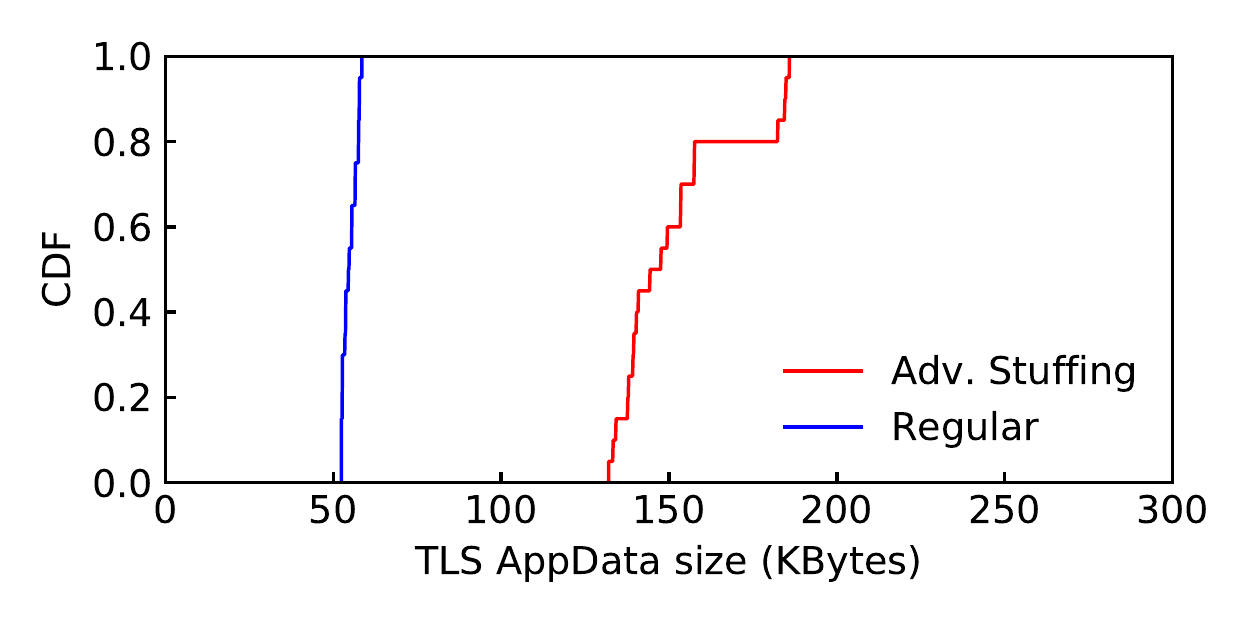}
    \caption{CDF of the total TLS AppData record byte count per sequence of 12 Metasploit commands.}
    \label{fig:tlsbytes}
\end{figure}

\begin{figure}[h!]
    \centering
    \includegraphics[width=0.9\linewidth,keepaspectratio]{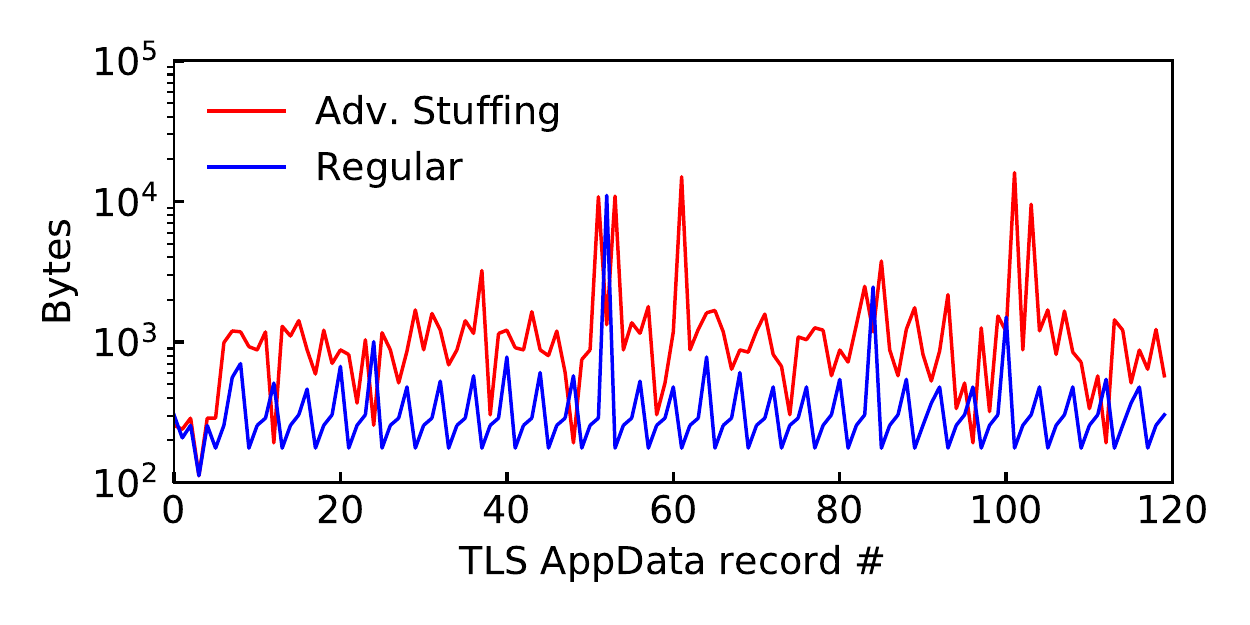}
    \caption{Sequence of TLS record sizes for Regular Metasploit vs. Metasploit with Adversarial Stuffing}
    \label{fig:OverheadSizes}
\end{figure}

We observe no significant changes in the runtime of a sequence of 12 Metasploit commands\footnote{\texttt{sysinfo, ps, getuid, getpid, ipconfig, route, pwd, ls, cat /etc/shadow, download /etc/shadow, webcam\_list, exit
}} 
and a 3 fold increase in the number of TLS AppData record bytes. We did 20 runs of the 12 commands sequence and discarded 1 outlier, for both adversarial stuffing modification and without modifications.
The resulting average runtime for the 12 command sequence is 16.43 s (with less than 0.01 s difference between different runs) for both regular and adversarial stuffing.
In addition to the 3 fold increase in byte count, 
Figure \ref{fig:tlsbytes} shows much more variation in the byte count values with the adversarial stuffing modification. This is consistent with adversarial stuffing introducing not only more data but also more random data in the communication. Figure \ref{fig:OverheadSizes} illustrates this with an example of comparing TLS AppData record sizes for this sequence  with and without our modifications.

\begin{figure}[h!]
    \centering
    \includegraphics[width=0.9\linewidth,keepaspectratio]{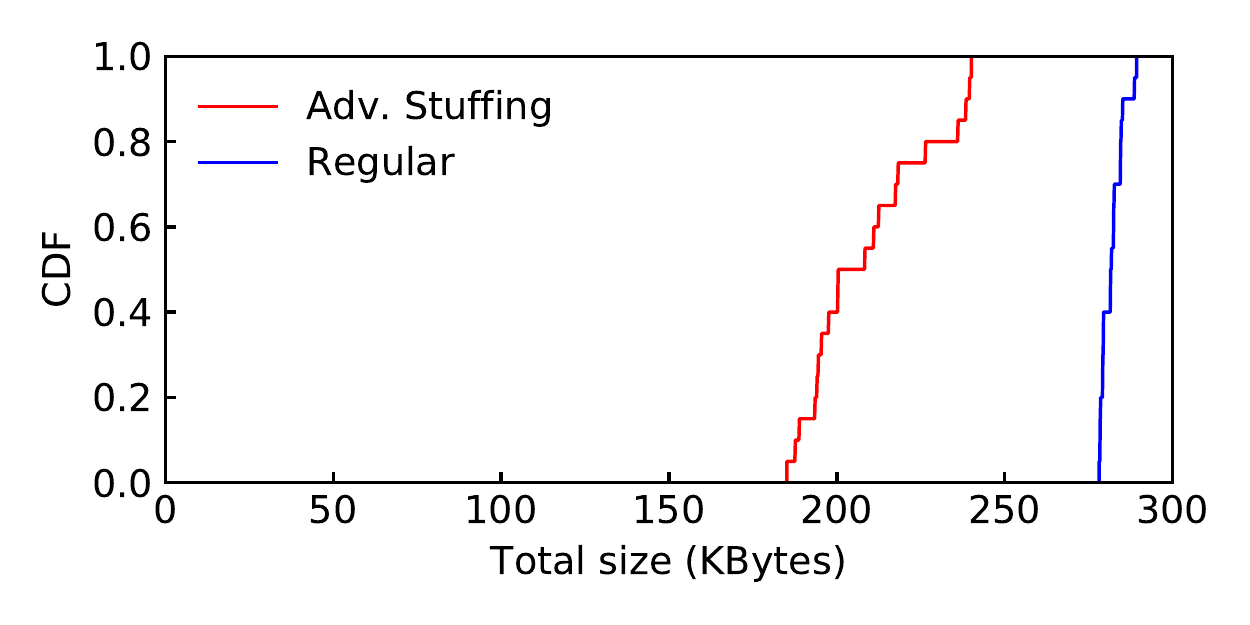}
    \caption{CDF of the total byte counts per sequence of 12 Metasploit commands.}
    \label{fig:pcapbytes}
\end{figure}

We also found that, unlike the number of TLS AppData record bytes, the total number of bytes sent between the framework and the payload is 25\% smaller for adversarial stuffing. Figure \ref{fig:pcapbytes} shows the distribution for total number of bytes. Although this sounds counter intuitive as we know we have more TLS AppData record bytes, it can be explained by the overhead of the TLS handshake for each TLS/TCP connection. In fact, the \textit{Multiple HTTP Request per TCP connection} modification reduces the number of TCP connections for the same number of Metasploit commands and, as such, reduces the TLS handshake overhead.

\FloatBarrier

\FloatBarrier

\FloatBarrier

\vspace{5mm}

\begin{figure}[t!]
    \centering
    \includegraphics[width=0.9\linewidth,keepaspectratio]{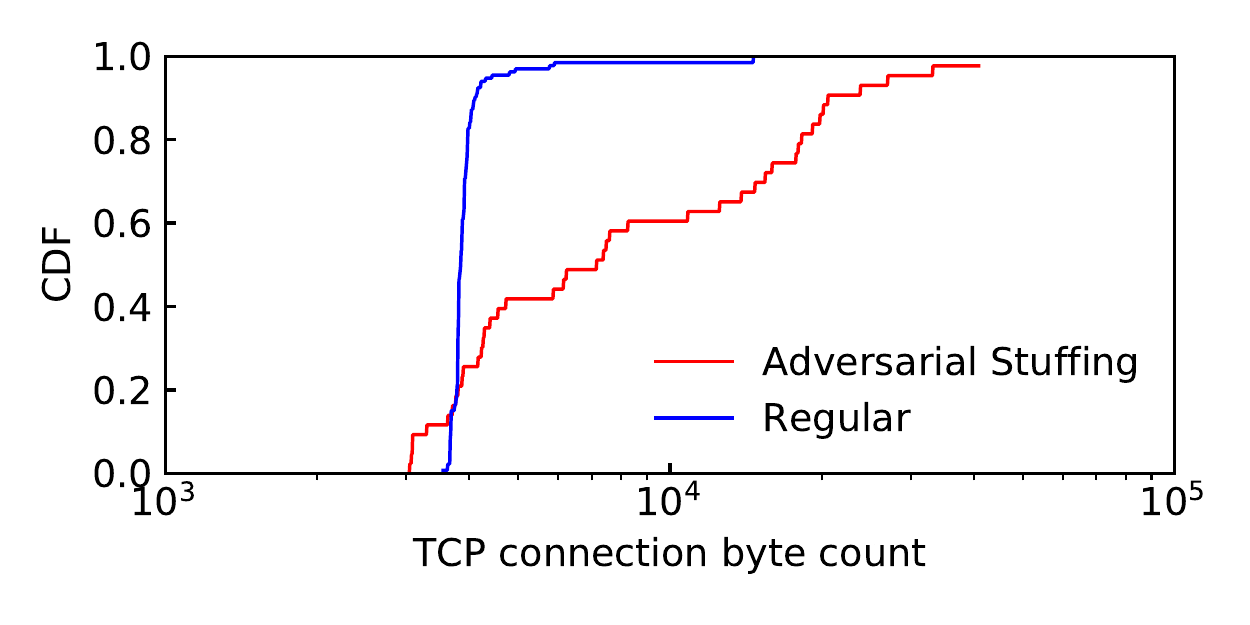}
    \includegraphics[width=0.9\linewidth,keepaspectratio]{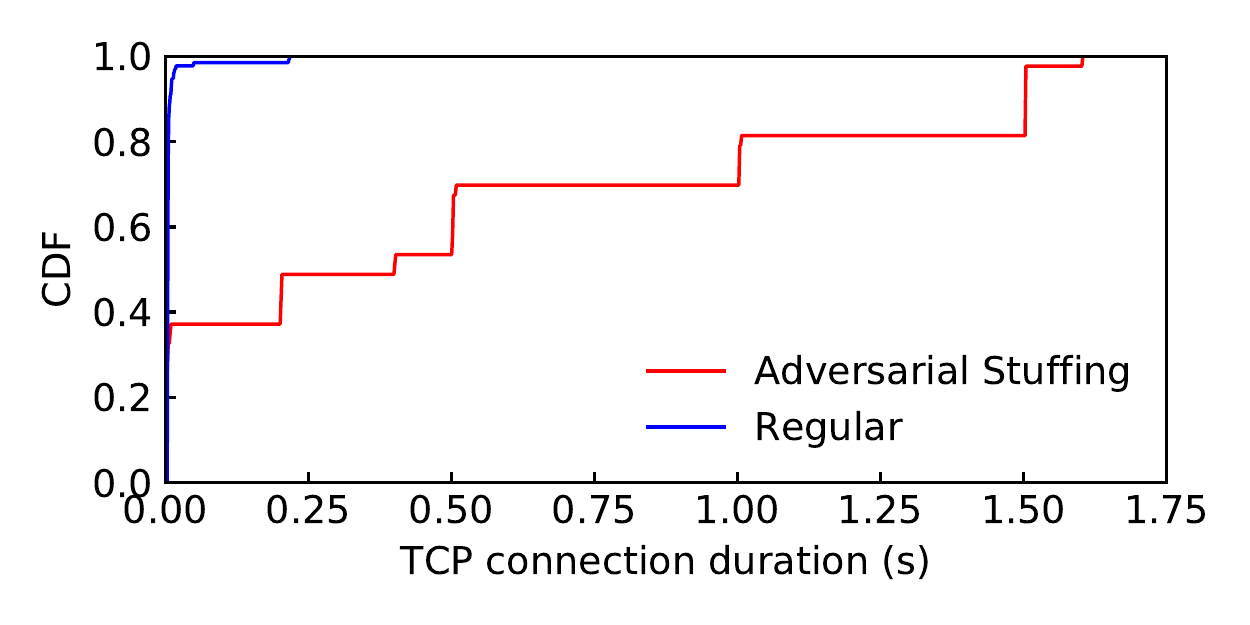}
    \caption{CDF of the byte counts and duration per TCP connection for an example sequence of 12 Metasploit commands.}
    \label{fig:oh-per-tcp-conn}
\end{figure}

Figure \ref{fig:oh-per-tcp-conn} (top) shows that the minimum byte count value for a TCP connection is between 3 and 4 KBytes, which mostly accounts for the handshake of the TLS connection including certificates. Most regular TCP connections (without our modifications) have close to the minimum value, which indicates that most of these connections are likely polling requests without any Metasploit commands. Figure \ref{fig:oh-per-tcp-conn} (bottom) also shows that regular TCP connections are mostly short lived, which is consistent with the polling pattern. The byte count CDF (Figure \ref{fig:oh-per-tcp-conn}, top) shows adversarial stuffing TCP connections have many different byte counts, as expected given they are trying to change the classifier decision by adding more random-sized stuffing and making the TCP connections last longer.
This is confirmed by the TCP connection duration CDF. Although most adversarial stuffing TCP connections have more bytes than regular connections, the total number of adversarial stuffing connections is approximately 4 times smaller than in the regular case, which compensates the increase of byte count per connection.

\begin{figure}[t!]
    \centering
    \includegraphics[width=0.9\linewidth,keepaspectratio]{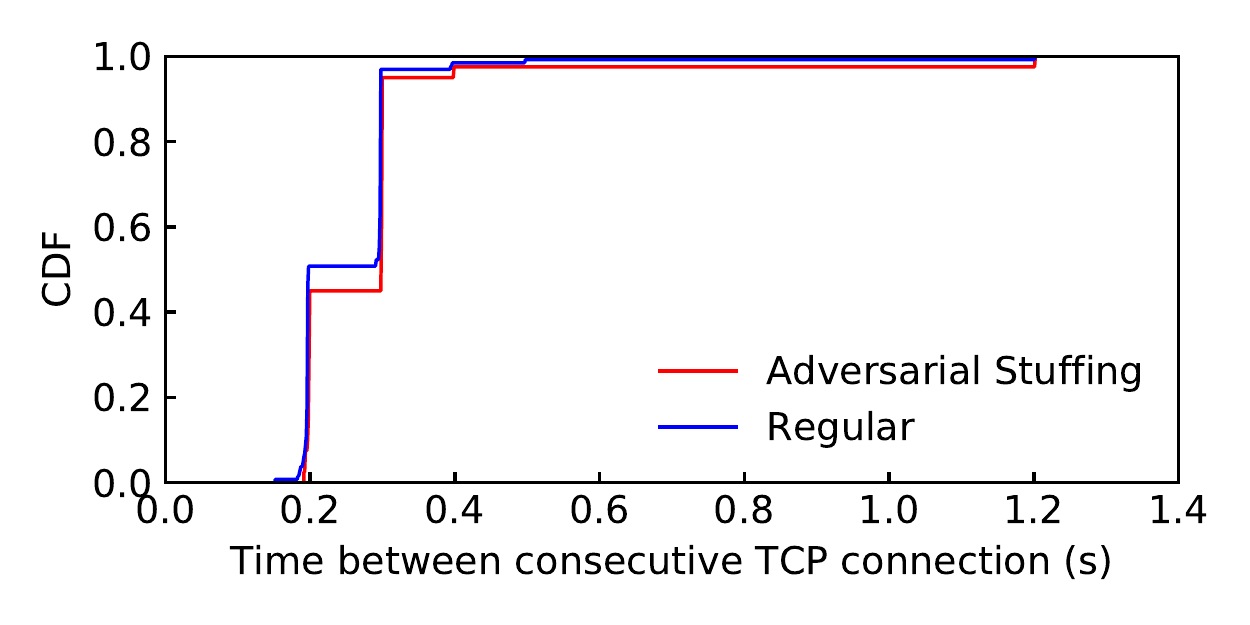}
    \caption{CDF of the time between consecutive TCP connections, for an example sequence of 12 Metasploit commands.}
    \label{fig:oh-per-tcp-conn-itf}
\end{figure}

We can also observe from Figure \ref{fig:oh-per-tcp-conn-itf} that the time between the end of a TCP flow and the beginning of the next flow is extremely similar when comparing the regular and adversarial stuffing schemes. In the regular communication scheme this represents the off period between two polling requests, and is likely related to the backoff that the payload introduces in the polling period when there are no Metasploit commands from the framework. We do not modify the original polling backoff mechanisms, although it would make sense if we were to evade a detector that uses the time between consecutive TCP flows as a feature.




\FloatBarrier

\section{Conclusion}

In this paper we validated what we suspect is the widely perceived notion that Metasploit C2 traffic can be easily detected even if it is wrapped under TLS encryption. To do so we trained a machine learning detector to distinguish between Metasploit TCP connections and web browsing TCP connections and validated this detector against Metasploit C2 traffic and web browsing traffic that we captured. This yielded 99\% accuracy on a balanced dataset. 
We then modified the Metasploit and Mettle source code so that its C2 traffic can evade the detector and showed that with two different modifications to the traffic -- 1) adding bytes, 2) grouping two or more Metasploit HTTP requests into a single TCP connection -- a pentester can evade the detector in almost 100\% of the Metasploit's TCP connections. We then retrained the detector with samples from the modified Metasploit C2 traffic, which again yielded a high (98.4\%) accuracy, and devised an approach to evade the new detector. This approach uses adversarial learning to choose how many bytes and TLS records a Metasploit C2 TLS connection should have in order to evade the improved detector. We conclude that the evasion rate drops slightly to 91\% when compared to the evasion rate of the initial detector and that if the modifications to the traffic are done only on the payload side then the evasion rate drops significantly to 50\%. We then look at the overhead of implementing the adversarial learning approach and find that although the size of the TLS payloads increases three fold due to our adding of bytes, the overall number of bytes sent between the Metasploit framework and the Mettle payload reduces by an average of 25\% due to a smaller overhead in the total number of TLS handshakes. We also observe that the runtime of Metasploit commands in the payload with our modifications does not increase when compared to the original Metasploit.

\FloatBarrier

In the future we intend to improve our Metasploit and Mettle modifications to split TLS records in two or more, thus considering the case where adversarial learning suggests a smaller TLS record size value than the one that needs to be sent. We also intend to choose adversarial samples according to the command that the Metasploit framework is about to issue to the payload rather than randomly choosing from a set of Metasploit samples. Regarding the adversarial learning approach, we expect to further study the cycle of retraining and adversarial learning; in this paper we did a single iteration of adversarial learning, but the detector could easily apply adversarial learning to its own model and retrain the model with the adversarial samples, which would lead to a cycle. We would like to understand the properties of such a cycle and whether after some point the detector or the pentester could prevail over the other. Finally, we expect to be able to study an alternative to adversarial learning that is closer to mimicry and attempts to squeeze Metasploit traffic into sequences of TLS record sizes obtained from non-Metasploit traffic, namely web.

We would not like to conclude this paper without providing some considerations about how the Metasploit modifications we study here could be used improperly -- and by improperly we mean outside of the context of whitehat pentesting such as those targeting third-party systems and which can be detected by honeypot studies~\cite{metasploit-study}. These modifications are relatively straightforward to implement given modest know-how in programming, networking, and the Metasploit framework, and we believe any modest threat actor could develop them. This is one reason for publishing this work -- so network defenders can better understand what happens when a modified Metasploit payload is deployed in their networks. Less mature cybercriminals -- whose actions likely generate the bulk of attacks these days -- and that have only user-level experience of Metasploit, may find it harder to develop these modifications. These users would likely resort to downloading the code from a repository. Because of that, we decided not to publish our Metasploit and Mettle source code.

\bibliographystyle{IEEEtran}
\bibliography{refs}

\vspace{12pt}

\end{document}